\documentclass[journal=jacsat,manuscript=article]{achemso}

\usepackage[version=3]{mhchem} % Formula subscripts using \ce{}
\usepackage{amssymb}
\usepackage{hyperref}

\author{Joseph D. Clark}
\affiliation{School of Molecular and Cellular Biology, University of Illinois at Urbana-Champaign, Urbana, IL 61801, USA}
\altaffiliation{These authors contributed equally to this work.}

\author{Tanner J. Dean}
\affiliation{Center for Biophysics and Quantitative Biology, University of Illinois at Urbana-Champaign, Urbana, IL 61801, USA}
\altaffiliation{These authors contributed equally to this work.}

\author{Diwakar Shukla*}
\affiliation{Center for Biophysics and Quantitative Biology, University of Illinois at Urbana-Champaign, Urbana, IL 61801, USA}
\alsoaffiliation{Department of Chemical and Biomolecular Engineering, University of Illinois at Urbana-Champaign, Urbana, IL 61801, USA}
\alsoaffiliation{Department of Bioengineering, University of Illinois at Urbana-Champaign, Urbana, IL 61801, USA}
\alsoaffiliation{Department of Chemistry, University of Illinois at Urbana-Chamapaign, Urbana, IL 61801, USA}
\email{diwakar@illinois.edu}

\title{Two for the Price of One: Integrating Large Language Models to Learn Biophysical Interactions}
\abbreviations{IR,NMR,UV}

\begin{document}

% \begin{tocentry}

% Some journals require a graphical entry for the Table of Contents.
% This should be laid out ``print ready'' so that the sizing of the

% Inside the \texttt{tocentry} environment, the font used is Helvetica
% 8\,pt, as required by \emph{Journal of the American Chemical

% The surrounding frame is 9\,cm by 3.5\,cm, which is the maximum
% permitted for  \emph{Journal of the American Chemical Society}
% graphical table of content entries. The box will not resize if the
% content is too big: instead it will overflow the edge of the box.

% This box and the associated title will always be printed on a
% separate page at the end of the document.

% \end{tocentry}

\begin{abstract}
Deep learning models have become fundamental tools in drug design. In particular, large language models trained on biochemical sequences learn feature vectors that guide drug discovery through virtual screening. However, such models do not capture the molecular interactions important for binding affinity and specificity. Therefore, there is a need to `compose' representations from distinct biological modalities to effectively represent molecular complexes. We present an overview of the methods to combine molecular representations and propose that future work should balance computational efficiency and expressiveness. Specifically, we argue that improvements in both speed and accuracy are possible by learning to merge the representations from internal layers of domain specific biological language models. We demonstrate that `composing' biochemical language models performs similar or better than standard methods representing molecular interactions despite having significantly fewer features. Finally, we discuss recent methods for interpreting and democratizing large language models that could aid the development of interaction aware foundation models for biology, as well as their shortcomings.
\end{abstract}
% by learning merged representations of the the internal layers..e
\subsection{Introduction}
\begin{figure}
    \centering
    \includegraphics[width=0.82\linewidth]{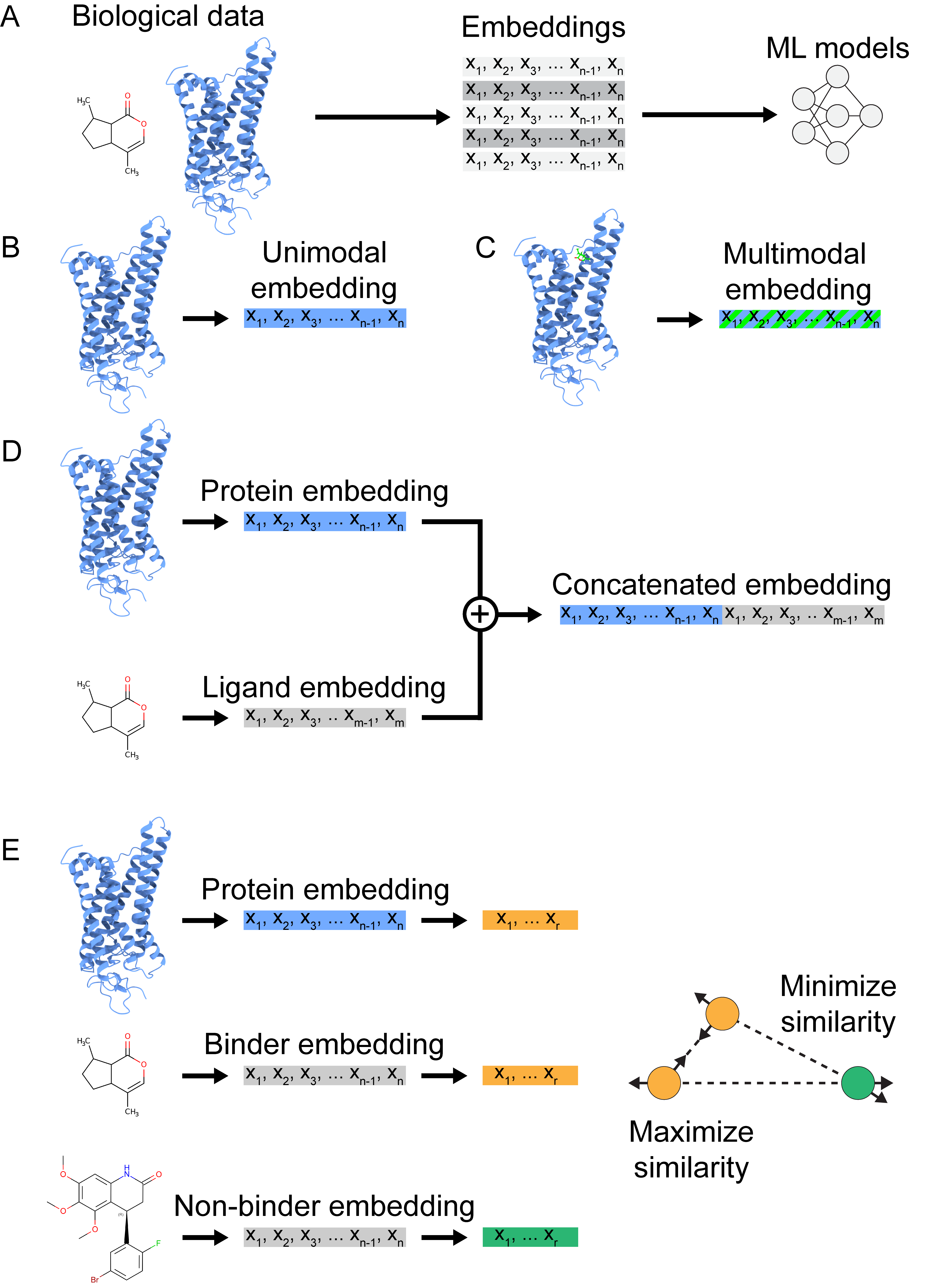}
    \caption{The workflow for computational screening and the development of multimodal embeddings. A) Sequence or structure data are mapped to feature vectors called embeddings and are used to train machine learning models to predict molecular properties. B) Unimodal embeddings contain information about a single molecule. C) Multimodal embeddings encode the features of molecular complexes. D) Concatenation appends two individual molecular embeddings to produce a larger multimodal representation. E) Contrastive learning projects two or more representations to a shared embedding space. The embeddings of interacting molecules are optimized to be closer in the latent space.}
    \label{fig:1}
\end{figure}

The vastness of chemical space severely limits experimental screening in drug design\cite{Reymond2010}. Advances in deep learning can help circumvent this issue by enabling large scale computational screening to identify potential drugs\cite{Sadybekov2023, Wu2024}. First, molecules are transformed into feature vectors (also called embeddings) which encode biochemical information (Figure \ref{fig:1}A). In practice, embeddings can take the form of hand picked features\cite{Wei2019}, topological encodings\cite{Rogers2010}, or the internal representations from large language models\cite{Lin2023}.  Machine learning (ML) models are then trained to predict molecular properties given embeddings as input. Popular ML models in drug discovery include random forests, support vector machines, and neural networks\cite{rodriguez2022evolution, kapsiani2021random, krishnan2021accelerating}. Accurate ML models enable efficient screening of molecular libraries to identify candidate molecules with desired properties. In the context of drug design, essential properties of interest include high binding affinity and specificity for protein target(s)\cite{Puszkarska2024, AbbasiMesrabadi2023}. Therefore, ML models often predict properties of molecular complexes, such as protein-ligand, protein-protein, protein-peptide, or protein-nucleic acid interactions. However, molecular representations are typically unimodal in nature and lack explicit features describing intermolecular interactions (Figure \ref{fig:1}B). Additionally, most molecular representations are derived from sequence alone, which limits their ability to capture structural information important for binding. Finally, large language models containing up to billions of parameters produce high-dimensional embeddings with uninterpretable features that cannot be easily mapped to known biochemical concepts. 

Failure to represent molecular interactions significantly reduces the performance of ML models trained on binding affinity and specificity. One solution is to join, merge, or `compose' unimodal molecular representations to produce augmented embeddings of molecular complexes (Figure \ref{fig:1}C). We review a variety of methods to merge standalone molecular representations into multimodal embeddings, and argue that future work should emphasize the fusion of molecular language models to balance a trade off between computational efficiency and representation ability. We emphasize recent work in natural language processing which seeks to `compose' domain specific language models by learning to merge the representations from internal layers. We argue that these composition frameworks could become powerful tools to integrate information from different chemical modalities. We also take note of parameter efficient fine-tuning (PEFT) and language model interpretability methods that could advance the development of composed langauge models. Finally, training tasks such as pairwise contact prediction could help infuse structural information into multimodal embeddings, producing an emergent understanding of binding affinity and specificity. 

\subsection{Concatenation Appends Standalone Representations}
Specificity and binding affinity are properties of molecular complexes. Therefore, there is a need to merge features from distinct molecular modalities to effectively represent biological interactions. The simplest method to combine features from different modalities is to concatenate molecular representations. Let $h_a \in \mathbb{R}^m$ and $h_b \in \mathbb{R}^n$ be two vector representations for distinct molecules (e.g., a protein ligand pair). Concatenation appends the two representations to produce a new encoding $h_{a}h_b \in \mathbb{R}^{m+n}$ which contains features of both molecules (Figure \ref{fig:1}D). The concatenated vector is used as the input to machine learning models that predict molecular interactions and their structural or kinetic properties.   

Numerous studies combine molecular representations through concatenation\cite{Kroll2023-fm, ztrk2018, Zeng2019, Nguyen2020, Jiang2020, AbbasiMesrabadi2023}. In Enzyme Substrate Prediction (ESP)\cite{Kroll2023-fm}, protein embeddings were obtained from the protein language model ESM-1b\cite{Rives2021} and small molecule embeddings were encoded via Graph Neural Networks (GNNs). The concatenated embeddings were used to train a gradient boosting decision tree to predict enzyme specificity. DeepDTA\cite{ztrk2018} employed a similar method to predict the binding affinity of protein-ligand pairs. Two independent convolutional neural networks were trained to produce feature vectors for proteins and ligands, then a multi-layer perceptron was trained on the concatenated representations to regress binding affinity. Other studies have explored concatenation for other modalities, such as DeepLigand\cite{Zeng2019} which concatenated Major Histocompatability Complex (MHC) class I protein feature vectors with peptide sequence representations to predict specificity. In this case, amino acid sequences were represented as one-hot encodings and Blosum50 matrix scores, and a deep residual network was trained on the conactenated embeddings.  The widespread use of concatenation is largely attributed to its role as a benchmark to compare more complex embedding strategies. Concatenation is a simple and parameter free strategy that supports the combination of arbitrary features from sources such as protein or chemical language models, hand crafted or biochemical features, and traditional molecular representations such as ECFPs. However, concatenated embeddings are usually high-dimensional, potentially leading to poorer performance on small data sets, overfitting, and longer training times for ML models that scale unfavorably with the number of input features. Lastly, appending the standalone representations of molecules does not explicitly encode their interactions which significantly limits the expressive power of concatenation. 

\subsection{Contrastive Learning Aligns Independent Embedding Spaces}
Contrastive learning encompasses a variety of related methods originally proposed for representation learning in computer vision\cite{LeKhac2020}. Broadly, contrastive learning trains embeddings by maximizing the similarity between encoded data points with shared properties (e.g., increasing the similarity of embeddings of ligands from the same group of agonists for a known protein). Importantly, separate encoders can be employed to perform multimodal contrastive learning in which similarity is optimized between embeddings of data points from distinct modalities (Figure \ref{fig:1}E)\cite{https://doi.org/10.48550/arxiv.2104.12836}. In the context of biology, contrastive learning models often maximize the agreement between learned embeddings of interacting molecules such as protein ligand pairs\cite{Singh2023}. Specifically, a protein target and one or more ligands are featurized by encoders such as pretrained language models or fingerprinting methods. In most cases, the encoder models are not trained during contrastive learning. Each molecular representation is then projected to a shared latent space by learnable transformations such as linear projections or feed-forward layers. Finally, a loss is calculated based on the similarity of pairs, triplets, or batches of embeddings in the shared latent space. In this framework, the probability of binding is predicted by the similarity/distance of protein and ligand embeddings in the shared latent space. The shared embedding space is structured to encode binding specificity and is typically of lower dimensionality than the original embeddings. Protein targets are usually termed as anchors, while binding and non-binding molecules are denoted as positive and negative samples respectively. Cosine similarity is typically used as the distance metric, and there are multiple related contrastive losses \cite{1467314, https://doi.org/10.48550/arxiv.1503.03832, https://doi.org/10.48550/arxiv.1511.06452}. The triplet loss minimizes the euclidean distance between one anchor and one positive while maximizing the distance between the anchor and a negative \cite{https://doi.org/10.48550/arxiv.1503.03832}. Other contrastive losses such as InfoNCE\cite{https://doi.org/10.48550/arxiv.1807.03748} and NT-Xent\cite{https://doi.org/10.48550/arxiv.2002.05709} stabilize training by expanding the triplet loss to include more negative samples. 

A major advantage of contrastive learning is the inference speed of the trained model due to the inexpensive nature of the cosine similarity computation. Once a large number of ligand and protein embeddings are precomputed, all pairwise interactions can be efficiently calculated as a cosine similarity matrix. Therefore, contrastive learning has the potential to dramatically expedite computational screening. However, contrastive learning may perform poorly on out-of-distribution predictions involving unseen targets that were not present as anchors in the shared embedding space\cite{Singh2023}. Despite this, multiple studies have successfully used contrastive learning models to predict drug-target interactions \cite{Singh2023, Gorantla2024, https://doi.org/10.48550/arxiv.2310.06367, xu2024multimodal, Du2024}. 

In ConPLex\cite{Singh2023}, protein representations were extracted from the language model ProtBert\cite{Brandes2022}, and small molecules were represented as Morgan Fingerprints. The embeddings were projected to a shared space via a learnable fully connected layers, and the model was jointly trained via the triplet loss and binary classification of interacting protein-ligand pairs. The probability of a drug-target interaction was interpreted as the sigmoid of the cosine similarity between protein and ligand embeddings. Similarly, BALM\cite{Gorantla2024} performed contrastive learning of protein-ligand binding affinity, but replaced ProtBert and Morgan Fingerprints with the language models ESM-2\cite{Lin2023} and ChemBERTa-2\cite{https://doi.org/10.48550/arxiv.2209.01712} respectively. A parameter efficient fine-tuning (PEFT) method was used to update the encoders during contrastive training. PEFT methods add a small number of trainable parameters to an otherwise fixed language model to learn minor modifications to the embeddings. Using PEFT to update the encoder models during contrastive training markedly improved binding affinity prediction. We provide a more detailed discussion of PEFT in a later section.

DrugCLIP\cite{https://doi.org/10.48550/arxiv.2310.06367} predicted interacting protein-ligand pairs via contrastive learning utilizing structure-aware ligand and protein binding pocket encoders.
In a later study, DrugCLIP was expanded to explore the scalability of contrastive virtual screening by predicting over 10 trillion protein-ligand interactions\cite{Jia2024}. Finally, PepPrCLIP\cite{Bhat2025} is a contrastive learning framework tailored for peptide binder design. Binding peptide-protein pairs were each encoded by ESM-2, and a shared latent space was learned via binary classification based on the cosine similarity of projected embeddings. During inference, Gaussian noise was added to the ESM-2 embeddings of known peptide binders, and the perturbed embeddings were decoded into novel sequences. Candidate sequences were ranked by their predicted binding to select for high affinity binders. While most studies use contrastive learning for prediction and design of molecular interactions, other works align protein sequence embeddings with structure representations\cite{Wang20242} or biophysical features\cite{Peng2025}.   

Contrastive learning models produce embeddings that implicitly capture binding specificity though distance in the latent space. The flexibility of contrastive learning supports the alignment of multiple modalities and fast inference speeds. However, such models still lack an explicit model of molecular interactions, suggesting a need for more sophisticated architectures to learn multimodal molecular representations.

\subsection{Attention Learns Interactions via Dynamic Reweighting}
\begin{figure}
    \centering
    \includegraphics[width=0.95\linewidth]{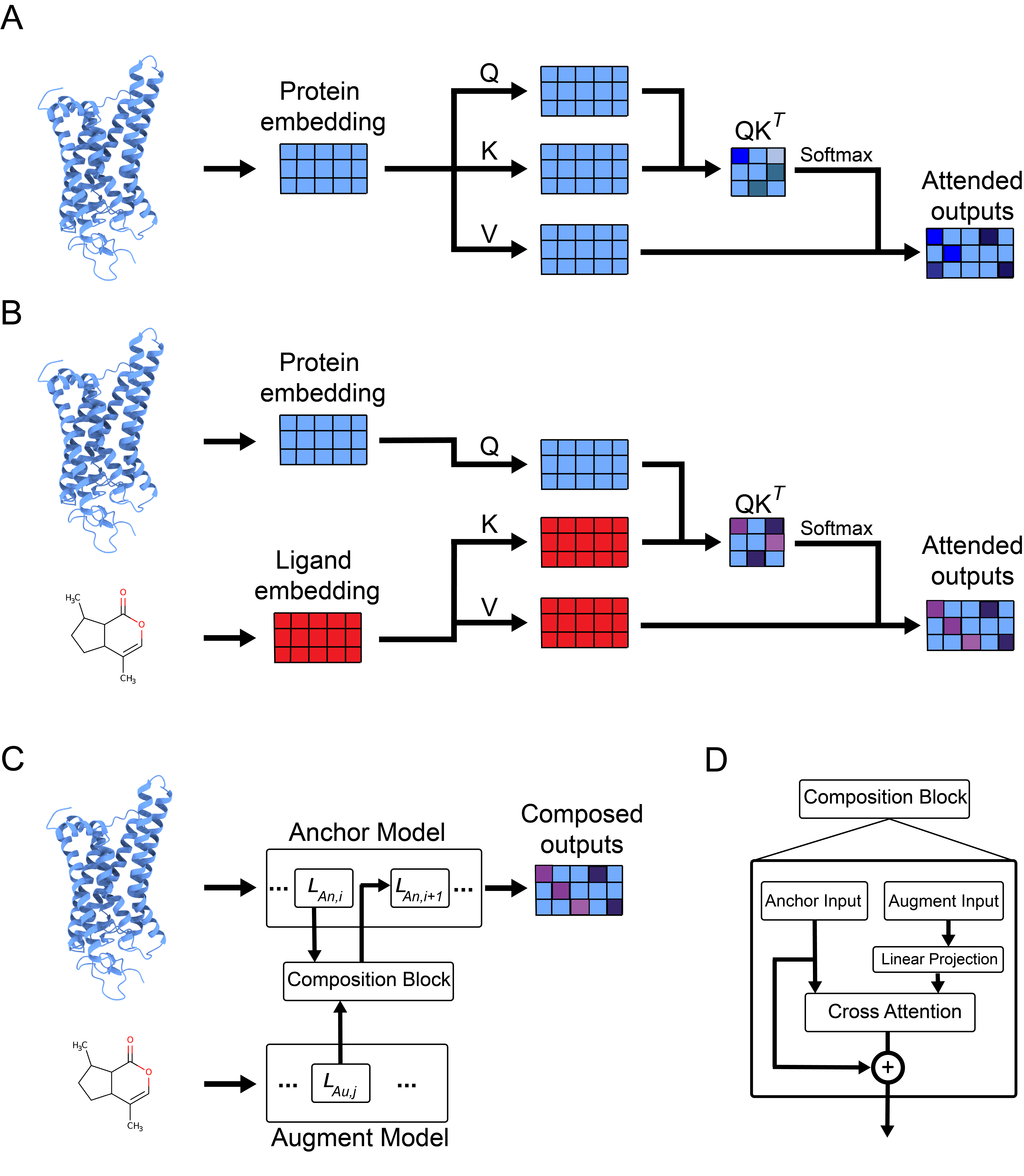}
    \caption{Schematic view of more complex methods of augmentation. A) The self-attention mechanism whereby an input tensor learns to place additional attention on specific elements of the sequence to dynamically focus on relevant parts of the input. B) The cross-attention mechanism, where a representative protein embedding acquires additional attention to key residues via a ligand embedding. C) Composition of two language models allows for the generation of ouptput embeddings of the anchor models dimensionality containing the relevant information of all augment models. D) The composition block of the CALM method uses a linear projection to match embedding dimensionalities prior to cross attention between embeddings and a skip connection into the next layer of the anchor model.}
    \label{fig:2}
\end{figure}

In the early days of neural machine translation, state of the art methods consisted of recurrent neural network (RNN) encoder-decoder architectures. While these architectures performed well, a major problem was the compression of information into a fixed-length vector regardless of the length of the input sentence. These early models tended to perform poorly on sentences which extended beyond the maximum length in the training corpus. Proposed in 2014, the initial goal of the attention mechanism was to improve translation accuracy by allowing the model to focus on specific parts of the input sequence relevant to each output token \cite{bahdanau2014}. Unlike prior methods, the attention mechanism enables the model to assign dynamic weights to each part of the input regardless of the sequence length. In the modern `self-attention' mechanism, embeddings are duplicated and transformed into `queries', `keys', and `values' by learned linear projections (Figure \ref{fig:2}A). Queries and keys are multiplied and scaled to produce a square attention matrix describing the importance of each position in the sequence to each other position. The attention matrix is multiplied by the values to produce an updated representation in which the embedding for each position is a linear combination of all other embeddings in the sequence. This flexibility marked a turning point, allowing models to selectively prioritize information without relying on a rigid sequential structure like prior RNN architectures. Following in the success of the attention mechanism, the introduction of the Transformer model in 2017 revolutionized language models by incorporating self-attention as a core component\cite{vaswani2017}. 
% using scaled dot product attention as:
% \begin{equation}
% \text{Attention}(Q, K, V) = \text{softmax}\left(\frac{QK^\top}{\sqrt{d_k}}\right)V
% \label{eq:1}
% \end{equation}
% where the query, key, and value matrices ($Q$,$K$,$V$ respectively) have learnable weight matrices to learn the attention between tokens\cite{vaswani2017}. 
% The addition of the scaling factor as shown in Equation \ref{eq:1} is an integral part of the transformers performance as it stabilizes the gradients for optimization by reducing the magnitude of the dot products. 
This architecture enabled efficient handling of long sequences and facilitated scalability to larger datasets and models, leading to the recent popularity of large language models (LLMs) like ChatGPT, Claude, and Llama\cite{touvron2023}. Cross-attention mechanisms extend the concept of self-attention to connect information across different data modalities. Specifically, queries and key/value pairs come from distinct modalities allowing tokens from one modality (e.g., structure) to update tokens from another (e.g., residue embeddings), thus integrating complementary information (Figure \ref{fig:2}B). In more recent years, this idea has been extended to biochemical research where information can be gained on protein-ligand tasks through cross attention of protein and ligand embeddings.

Following the success of natural language processing, many biological and chemical problems have been composed as sequence problems such as the sequence of nucleotides in DNA/RNA, sequence of amino acids in a protein, and even the SMILES string of small molecule ligands to take advantage of cross-attention's ability to connect information across sequence, structure, and biochemical features. This ability to connect information across different data has played a critical role in many recent applications of LLMs to biochemistry. For example, ChemGLaM used a cross-attention ``interaction block" between the chemical language model MolFormer, and the protein language model ESM2 to improve compound-protein interaction (CPI) predictions \cite{Koyama2024, Ross2022, Lin2023}. These interaction-aware embeddings were then concatenated to ESM-2 embeddings and fed into a final fully-connected network for CPI predictions. ChemGLaM consistently outperformed previous state of the art methods for sequence only CPI predictions across four benchmarks. Additionally, recent work on lasso peptides and their corresponding cyclases has led to the development of LassoESM \cite{Mi2024}. LassoESM uses cross attention between ESM2 and a fine-tuned language model for lasso peptides, enabling state-of-the-art performance in cyclase-peptide pair prediction along with other tasks such as the prediction of RNA polymerase inhibitory activity and non-cognate pairs of peptides and cyclases. Along with these prediction tasks, ablation studies conducted on LassoESM demonstrated the importance of cross attention over concatenation of embeddings for improved model performance. Similarly, the PepNN model leverages reciprocal multi-head cross attention between peptide sequence and protein structure information to predict peptide binding sites\cite{Abdin2022}. The model was pre-trained on protein-protein interfaces from the Protein Data Bank\cite{Berman2000} before being fine-tuned on protein-peptide complexes. Cross attention's ability to focus information between modalities enabled this architecture to achieve a strong performance of 0.88 ROC AUC on a per-residue binding score. Finally, CoNCISE\cite{Erden2025} learned hierarchical discrete representations of chemical space and performed cross-attention with protein embeddings to achieve computationally efficient drug-target interaction prediction. Interestingly, cross attention mechanisms are agnostic to type of cross-modal information sharing within models, and multiple works develop strucutre-aware protein embeddings via cross attention between structure and sequence representations\cite{Wang20242, Tan2024}. 
% , as these models continue to grow in size one must also consider how to manage the curse of dimensionality.

Attention mechanisms, more specifically cross attention, revolutionized the ability to share information across modalities. The method has allowed for the integration of protein structure and ligand information among others, enhancing the prediction of interactions and binding affinities, which is crucial for drug discovery and protein-ligand docking studies. While these methods do improve feature information for cross-modal tasks, they do not address the problem of dimensionality that arises from sharing information between several embeddings leading to reduced model interpretability, a higher risk of overfitting to noise in the features, and increased computational complexity.

\subsection{Composition Merges Language Models to Integrate Domain-Specific Knowledge}

Following attention, much focus has been placed on alternative methods to combine multimodal information while avoiding the complexity of additional features. This area, known as composition, focuses on the modularity of language models which enables them to be modified and combined into larger systems for resolving more complex tasks. Composition methods merge language models in parameter space or data-flow space\cite{Akiba2025}. Composition in parameter space involves integrating the weights of multiple pretrained models with a shared architecture. In the simplest case, model parameters are combined via linear interpolation or vector averaging\cite{https://doi.org/10.48550/arxiv.2203.05482}. Fisher-Weighted averaging\cite{Matena2021} is a more rigorous method to average model weights while retaining maximal performance on distinct tasks. Finally, task-arithmetic\cite{ilharco2023editing} computes `task vectors' defined as directions in parameter space representing the difference between pretrained and fine-tuned models. Combining the task vectors of multiple models enables multi-task capabilities. Composing models in the data flow space involves re-routing intermediate representations through distinct models\cite{Akiba2025}. For example, an intermediate language model embedding may pass through a layer from a different language model before being forwarded to the next layer.  

% Finally, mixture-of-experts frameworks\cite{Du2022}.
% %ESM-3\cite{hayes2024}. 
A recent method building off of these ideas is Composition to Augment Language Models (CALM)\cite{bansal2024llm}. Unlike prior methods such as concatenation or attention, CALM merges two or more pretrained language models via a joint training task to achieve high performance on tasks requiring information from all composed models (Figure \ref{fig:2}C). CALM includes a baseline model called an `anchor' and one or more `augment' models which infuse multimodal information by performing cross-attention between multiple internal layers. Composition blocks are composed of a linear projection layer to match the embedding dimensionalities, and a cross attention layer to combine information (Figure \ref{fig:2}D). The attended output is then fed back into the anchor model via a residual connection (i.e., a vector addition). Composition blocks are interspaced throughout the encoder layers of a given anchor model and all augment models. During training on a joint task, the model learns to compose embeddings from distinct language models into a joint latent space while retaining the information of individual models and improving performance on tasks that leverage information from both models. Importantly, CALM retains the feature length of the anchor model while adding new information from multiple pretrained language models. Previous tasks demonstrated to work with this method have been limited to non-biological problems such as key-value arithmetic and low-resource language inclusivity. As biochemical language models continue to increase in complexity, such as ESM-3 with up to 98 billion parameters\cite{Hayes2024}, CALM-style methods could enforce a balance of model performance and representation ability while retaining reduced feature scales for general applications. In particular, CALM frameworks applied to foundational chemical and biological language models could produce powerful multimodal encoders with minimal increase in computational cost. 

\subsection{Performance Comparison of Multimodal Embeddings}

We compared the performance of the four multimodal representation strategies and visualized their embeddings. We trained models to classify peptide binders of MHC Class I and II proteins using the data set from Motmaen \textit{et. al}\cite{Motmaen2023}. Protein and peptide sequences were first encoded by lightweight versions of ESM-2\cite{Lin2023}, and 4 model architectures were trained to predict MHC specificity. A 2-layer multi-layer perceptron (MLP) was trained on concatenated peptide-protein embeddings. A contrastive learning model used two independent MLPs to project peptide/protein embeddings to a shared latent space with a cosine similarity loss for training. A cross attention model performed cross attention between the final-layer representations from each of the language models. Finally, a composition style model performed cross attention between the intermediate layers of the language models. As a baseline, we also trained 2 MLPs on unimodal peptide and protein embeddings respectively. We trained each model using 3 seeds and reported the average performance on a held-out validation set containing unseen peptide-protein pairs. For each seed, all models were trained for 100 epochs with a learning rate of $1\times10^{-3}$ and a batch size of 128. 

\begin{figure}
    \centering
    \hspace*{-0.45in}\includegraphics[width=0.82\linewidth]{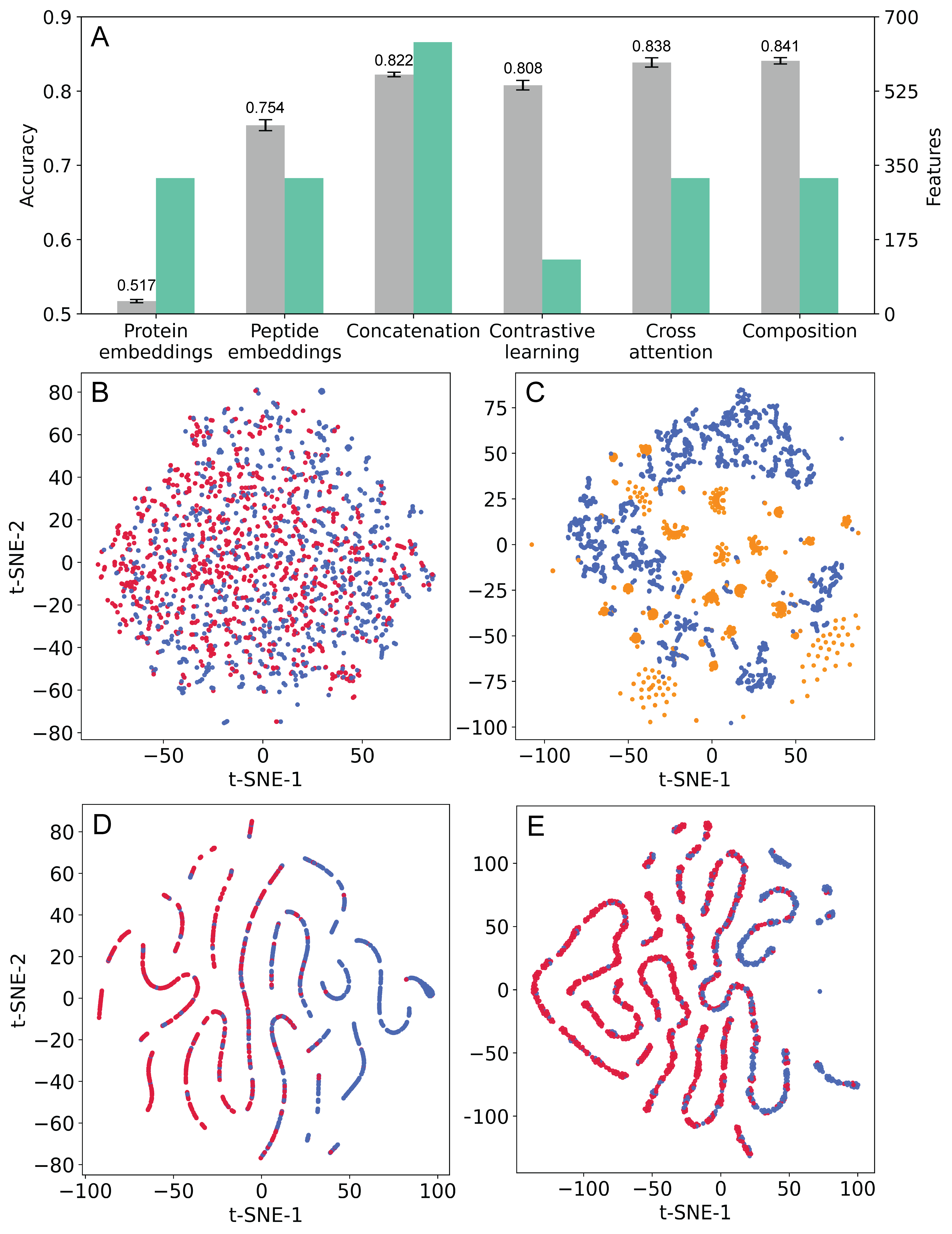}
    \caption{Comparison of multimodal embeddings on an MHC class I and II protein specificity prediction task. A) Cross attention and composed embeddings perform the best at peptide binding specificity prediction, with the embedding size shown on the right. B) Concatenated peptide-protein embeddings do not immediately distinguish between binding (red) and non-binding (blue) sequence pairs. C) Contrastive learning maps protein embeddings (orange) and peptide embeddings (blue) to a shared latent space. D) Cross attention and E) composition produce embedding spaces that cluster complexes by protein and separate binding/non-binding pairs.}
    \label{fig:3}
\end{figure}

Models trained solely on protein embeddings could not predict MHC specificity because a given protein may participate in multiple binding or non-binding pairs (Figure \ref{fig:3}A). In contrast, models trained on peptide embeddings showed moderate predictive power since all MHC proteins share some degree of specificity. Concatenated peptide-protein embeddings and contrastive learning further improved classification performance. Cross attention and composition models performed the best, suggesting that these embeddings learned the most information rich features describing peptide-protein interactions. The multimodal embeddings of peptide-protein pairs unseen during training were then reduced to 2 dimensions with t-distributed stochastic neighbor embedding (t-SNE) for visualization. Concatenated embeddings failed to separate binding and non-binding peptide-protein pairs prior to model training (Figure \ref{fig:3}B). The contrastive latent space grouped individual proteins into well-defined clusters based on shared specificity, and distributed peptide embeddings to ensure proximity to preferred targets (Figure \ref{fig:3}C). Finally, cross attention and composition style models learned embeddings of peptide-protein complexes  that clearly separated binding and non-binding pairs (Figure 
\ref{fig:3}D, E). Of particular note, these models also clustered complexes based on the identity of the protein (individual string-like clusters in Figure \ref{fig:3}E tend to contain similar proteins). In general, composition style models introduce minor perturbations to an original protein embedding based on the identity of a ligand, tailoring the protein embedding to context of a multimeric complex. We view composed embeddings as particularly powerful given their ability to simultaneously capture specificity while retaining valuable information about protein sequence identity.

\subsection{Improving the Accessibility and Transparency of Composed Models}
\begin{figure}
    \centering    \includegraphics[width=0.9\linewidth]{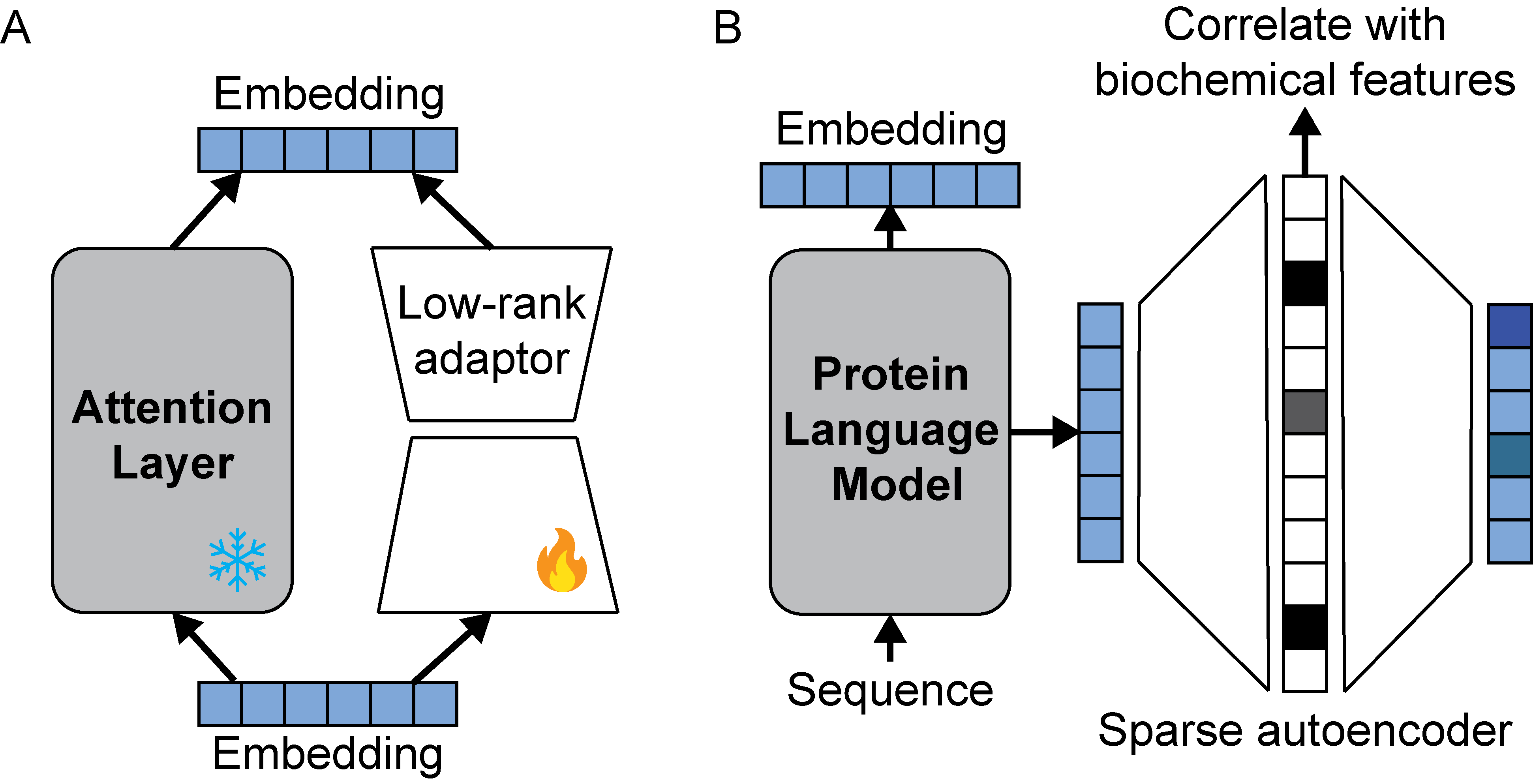}
    \caption{A schematic representation of large language model interpretability and parameter efficient fine-tuning (PEFT). A) Using sparse autoencoders, embeddings are projected to high dimensional, sparse representations which can be more easily interpreted. Sparse representations are used to reconstruct embeddings. B) In low rank adaptation (a representative PEFT method), language model representations are simultaneously modified by frozen pretrained weights, and `adaptors' containing a small number of learnable parameters. The output of a layer is slightly modified by the adaptors, enabling efficient model training.}
    \label{fig:4}
\end{figure}

The favorable properties of composed embeddings do not address the uninterpretable nature of language models or the significant computational resources needed to train them. Unfortunately, accurate language model predictions are less impactful if they do not offer direct biological insight or are inaccessible without high performance compute resources. Fortunately, recent methods from the broader natural language processing field may alleviate these issues by promoting cheaper and more transparent language model training. Parameter efficient fine-tuning (PEFT) seeks to train/fine-tune language models while updating few or none of the original parameters\cite{https://doi.org/10.48550/arxiv.2312.12148}. Instead, PEFT methods introduce lightweight `adapters' which learn to modify an existing language model's representations towards a distinct training objective. In Low-Rank Adaption (LoRA),\cite{https://doi.org/10.48550/arxiv.2106.09685} a language model's parameters are frozen and trainable adapters are added to the attention layers (Figure \ref{fig:4}A). An adapter projects the attention layer's input to a low-rank representation before reshaping it to the original dimensionality. The adapter's output is then added to the output of the pretrained attention layer. The low-rank projection of the input can contain as few as one or two dimensions, leading the adapter to contain considerably fewer parameters than the pretrained layer. LoRA enables dramatic improvements in training efficiency and leads to minimal performance degradation on fine-tuning tasks\cite{https://doi.org/10.48550/arxiv.2106.09685}. This has lead to widespread use of LoRA including during training of protein language models\cite{Sledzieski2024, Zhou2024, Schmirler2024}. Related PEFT methods find alternative modifications and decompositions of pretrained weights (see IA3\cite{https://doi.org/10.48550/arxiv.2205.05638} and DoRA\cite{https://doi.org/10.48550/arxiv.2402.09353}) or extend parameter efficient principles to language model pretraining (see GaLore\cite{https://doi.org/10.48550/arxiv.2403.03507}).

Language model interpretability remains an unsolved problem both in natural language processing and computational biology. Often, it is unclear what learned features enable models' powerful predictive performance. The high dimensionality and vast size of latent spaces make it impossible to parse interpretable features by directly analyzing molecular embeddings. Mechanistic interpretability provides insight into what a model has learned based on the assumption that language models represent more features than there are dimensions in their latent space\cite{bricken2023monosemanticity}. In this method, a `sparse-autoencoder' is trained to project language model embeddings to a substantially higher dimensional space (Figure \ref{fig:4}B). The model is trained to reconstruct protein embeddings using a linear projection. Importantly, the high-dimensional projection is explicitly trained to be sparse (i.e., most values are zero). Given a sparse protein representation, the few active features are assumed to correspond to biochemical properties that describe the corresponding sequence. Mechanistic interpretability has identified features that are only active in the presence of specific protein properties (e.g. zinc fingers, kinase binding sites)\cite{Simon2024}. This presents and exciting opportunity to shed light on what biochemical language models have learned. Of particular note, comparing the features learned by biochemical language models and domain specific or composition models could provide valuable insight on the specific biological concepts that emerged during training. 

\subsection{Future Perspective}
The development of methods to extract information from large language models—including contrastive learning, concatenation, attention mechanisms, and composition—has significantly enhanced their ability to process and generate complex language patterns. Each method offers unique advantages: contrastive learning improves model robustness through distinguishing relevant from irrelevant data; concatenation provides simplicity and flexibility by integrating diverse data streams; attention mechanisms enable efficient context capture and long-range dependencies; and composition promotes a structured and modular approach to knowledge representation. 

\begin{figure}
    \centering    \includegraphics[width=0.9\linewidth]{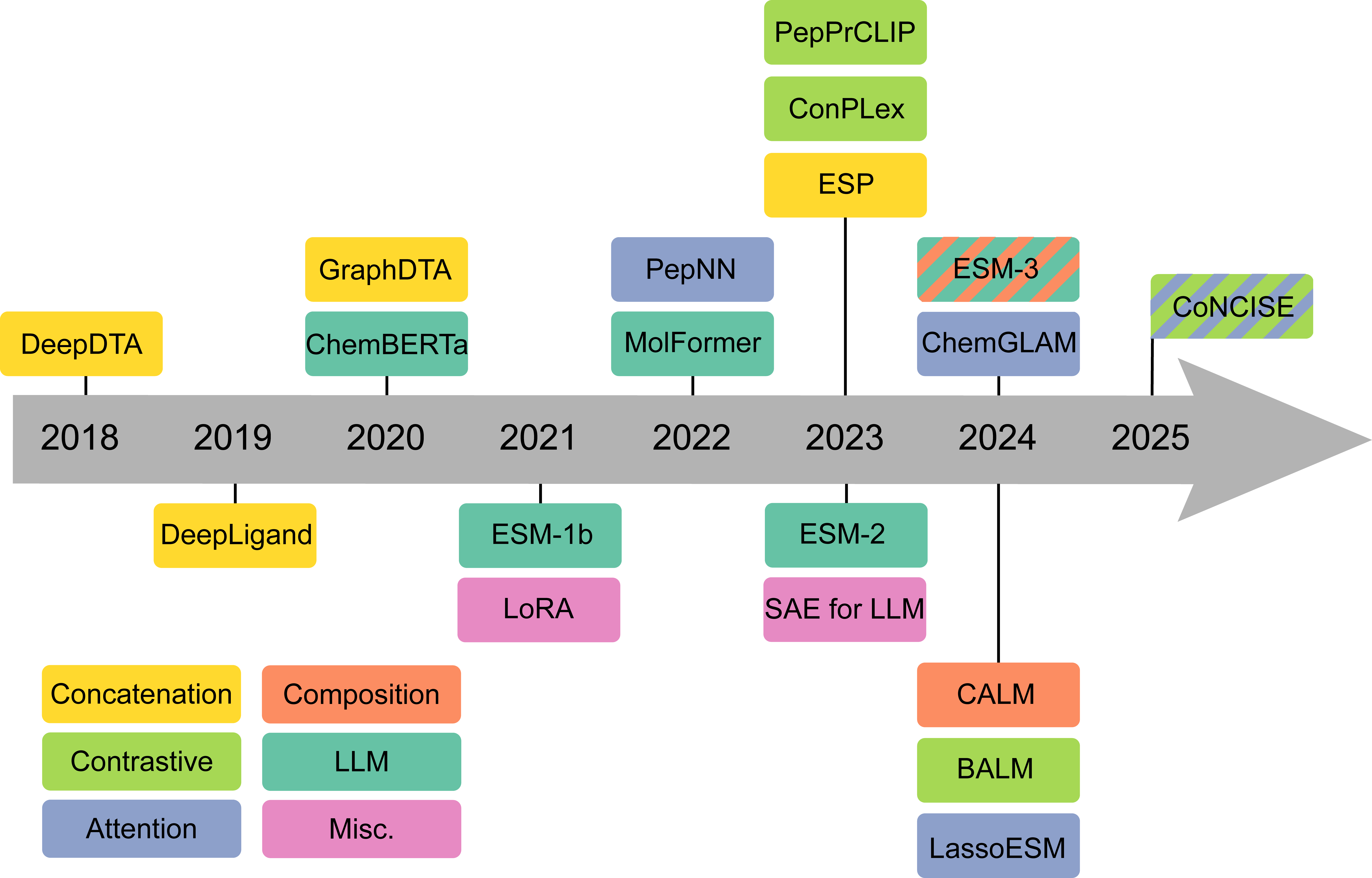}
    \caption{A timeline of deep learning methods to combine molecular representations. Methods are ordered by release date (e.g., as a preprint) rather than publication date. Attention and contrastive learning methods have grown in popularity along with hybrid methods.}
    \label{fig:5}
\end{figure}

Figure \ref{fig:5} presents a timeline of the major deep learning methods for merging multimodal embeddings. Contrastive learning and attention based methods have significantly grown in popularity, and methods that leverage multiple strategies have become more common. Moving forward, a key area for the development of language models lies in the integration of techniques to improve feature representations while evading the bottleneck of higher dimensions. Empirically, medium and small sized protein language models show competitive performance with larger models\cite{Vieira2024}. Recent work also suggests that protein language model representations can be compressed to significantly lower dimensionality with minimal loss of information needed for accurate structure prediction\cite{Lu2024a, Lu2024b}. This calls into question the idea that larger models and more features are needed to develop high performing, multi-modal embeddings. Further, expanding the size of language models presents a challenge to their interpretability as the question of how to explain their learned features is still outstanding\cite{https://doi.org/10.48550/arxiv.2402.01761}. However, recent work suggests that protein language model features can be correlated with biological concepts through `dictionary learning' methods called sparse auto encoders\cite{Simon2024}. The features of sparse autoencoders can be correlated with the biochemical properties, offering insight into what the model has learned. The application of interpretability methods to biology could help discover the emergent features learned by composed models. 

We propose that future work should prioritize using fewer, more salient features learned from interaction aware language models based on composition inspired frameworks. In particular, training CALM-like frameworks on the features of molecular complexes such as contact maps, binding affinities, or specificity could produce powerful multimodal encoders with structure aware embeddings. Inspired by PEFT methods, CALM frameworks promote efficiency during training, as only small composition blocks must be trained instead of entire language models. The nature of the CALM framework, coupled with recent work on domain-specific biological language models\cite{Zeng2024, Wang2024, Clark2024, Mi2024, Vincoff2025} makes it a promising technique for developing augmented language models of molecular interactions. 

% \begin{acknowledgement}

% Please use ``The authors thank \ldots'' rather than ``The
% authors would like to thank \ldots''.

% The author thanks Mats Dahlgren for version one of \textsf{achemso},
% and Donald Arseneau for the code taken from \textsf{cite} to move
% citations after punctuation. Many users have provided feedback on the
% class, which is reflected in all of the different demonstrations
% shown in this document.

% \end{acknowledgement}

% %%%%%%%%%%%%%%%%%%%%%%%%%%%%%%%%%%%%%%%%%%%%%%%%%%%%%%%%%%%%%%%%%%%%%
% %% The same is true for Supporting Information, which should use the
% %% suppinfo environment.
% %%%%%%%%%%%%%%%%%%%%%%%%%%%%%%%%%%%%%%%%%%%%%%%%%%%%%%%%%%%%%%%%%%%%%
% \begin{suppinfo}

% This will usually read something like: ``Experimental procedures and
% characterization data for all new compounds. The class will
% automatically add a sentence pointing to the information on-line:

% \end{suppinfo}

\bibliography{achemso-demo}

\end{document}